\documentclass[aps,showpacs,twocolumn,10pt,superscriptaddress,groupedaddress,floatfix]{revtex4-1}

\usepackage{bm}
\usepackage{graphicx}        
\usepackage{amssymb}
\usepackage{amsmath}
\usepackage{color}
\usepackage{titlesec}

\definecolor{seccolor}{RGB}{25,25,120}
\definecolor{subseccolor}{RGB}{20,90,15}
\definecolor{somtitlecolor}{RGB}{200,35,20}

\def\cephase{\varphi_{\text{CE}}}

\titleformat{\section}[hang]	
  {\bfseries\sffamily\Large\color{seccolor}} 
  {\color{seccolor}\thesection} 
  {1ex} 
  {} 
  [] 

\begin{document}

\title{Attosecond-recollision-controlled selective fragmentation of polyatomic molecules}

\author{Xinhua~Xie$^1$}
\author{Katharina~Doblhoff-Dier$^2$}
\author{Stefan~Roither$^{1}$}
\author{Markus~S.~Sch\"{o}ffler$^{1}$}
\author{Daniil~Kartashov$^1$}
\author{Huailiang~Xu$^{1,3}$}
\author{Tim~Rathje$^{4,5}$}
\author{Gerhard~G.~Paulus$^{4,5}$}
\author{Andrius~Baltu\v{s}ka$^1$}
\author{Stefanie~Gr\"{a}fe$^2$}
\author{Markus~Kitzler$^{1}$}
\email[Corresponding author: ]{markus.kitzler@tuwien.ac.at}

\affiliation{$^1$Photonics Institute, Vienna University of Technology, A-1040 Vienna, Austria, EU}
\affiliation{$^2$Institute for Theoretical Physics, Vienna University of Technology, A-1040 Vienna, Austria, EU}
\affiliation{$^3$State Key Laboratory on Integrated Optoelectronics, College of Electronic Science and Engineering,
Jilin University, Changchun 130012, China}
\affiliation{$^4$Institute of Optics and Quantum Electronics, Friedrich-Schiller-University Jena, D-07743 Jena, Germany, EU}
\affiliation{$^5$Helmholtz Institute Jena, D-07743 Jena, Germany, EU}

\begin{abstract}
Control over various fragmentation reactions of a series of polyatomic molecules (acetylene, ethylene, 1,3-butadiene) by the optical waveform of intense few-cycle laser pulses is demonstrated experimentally. 
We show both experimentally and theoretically that the responsible mechanism is inelastic ionization from inner-valence molecular orbitals by recolliding electron wavepackets, whose recollision energy in few-cycle ionizing laser pulses strongly depends on the optical waveform.
Our work demonstrates an efficient and selective way of pre-determining fragmentation and isomerization reactions in polyatomic molecules on sub-femtosecond time-scales.
\end{abstract}

\pacs{33.80.Rv, 42.50.Hz, 82.50.Nd}

\maketitle



Photodissociation of molecules by femtosecond light pulses is a fascinating phenomenon both in terms of underlying physics and practical implications to selective control of fragmentation pathway(s) through the shape of the optical waveform. Triggered by optical field ionization, the outcome of the complex electronic and subsequently nuclear dynamics leading to fragmentation is to a large extent predetermined on a sub-cycle time scale because of a sharp temporal localization of electron release to an attosecond interval at the peak of an intense optical half-cycle. Consequently, as has been shown in the case of diatomic molecules ---
D$_2$ \cite{Kling2006, Kling2008, Znakovskaya2012}, H$_2$ \cite{Kremer2009, Fischer2010}, HD \cite{Kling2008}, DCl \cite{Znakovskaya2011} and CO \cite{Znakovskaya2009, Liu2011} ---
dissociative ionization strongly depends on the carrier-envelope phase (CEP) of a few-cycle ionizing laser pulse.
The CEP-dependence in the case of the lightest molecules, D$_2$/H$_2$, has been linked to a field-driven population transfer between the binding ground state and the dissociative excited state. 

In this Letter, we present the first, to our knowledge, CEP control of fragmentation pathways of polyatomic molecules (acetylene, ethylene, 1,3-butadiene). Given the increased number of participating nuclei and a vastly more complex valence electron dynamics and the structure of energy surfaces, it is counterintuitive to expect the CEP-dependence of fragmentation pathways to survive in polyatomic molecules. Yet the experimental evidence provided in this Letter reveals a prominent role of the CEP for all three polyatomic species under examination and necessitates a search for a suitably controllable attosecond physical mechanism that persists despite the increased complexity of the system. We argue both experimentally and theoretically, that such a universal attosecond mechanism of quasi-single-cycle fragmentation of large molecules is the field control of the tunneled-out electronic wave packet that is steered by the optical waveform before its recollision with the parent ion. This scenario is indeed reminiscent to the role a light-field-accelerated electron plays in the process of high-harmonic generation \cite{corkum1993}. The universality of the CEP-effect in polyatomic molecular fragmentation is based on the fact that it is comparatively easy to ``encode'' the detached electron wave packet with a necessary recollision momentum. Conversely, it is difficult to encode many electronic degrees of freedom in a complex molecule directly with the electric field of the femtosecond pulse. 


In our experiments, we focus few-cycle laser pulses with sub-5 fs duration, linearly polarized along the $z$-direction of our lab coordinate system, in an ultra-high vacuum chamber onto a supersonic molecular gas jet, propagating along $x$, of randomly aligned acetylene (C$_2$H$_2$), ethylene (C$_2$H$_4$), or 1,3-butadiene (C$_4$H$_6$) molecules. Details on the experiment can be found in the Supplemental Material \cite{SOM}. In short, the three-dimensional momentum vectors of fragment ions created by laser-induced fragmentation of a single molecule were measured by cold target recoil ion momentum spectroscopy \cite{Doerner2000, Zhang2012}. 
The duration and the CEP of each few-cycle laser pulse was measured on a single-shot basis by stereo-detection of photoelectron spectra in a separate apparatus \cite{Wittmann2009, Sayler2011, Sayler2011a}, and linked to the momentum of the ionic fragments in the offline data analysis. 
The peak intensity of the laser pulses on target was determined from separate calibration measurements using single ionization of argon atoms in circularly polarized light \cite{Maharjan2005}.

\begin{figure}[!tb]
\centering
\includegraphics[width=\columnwidth]{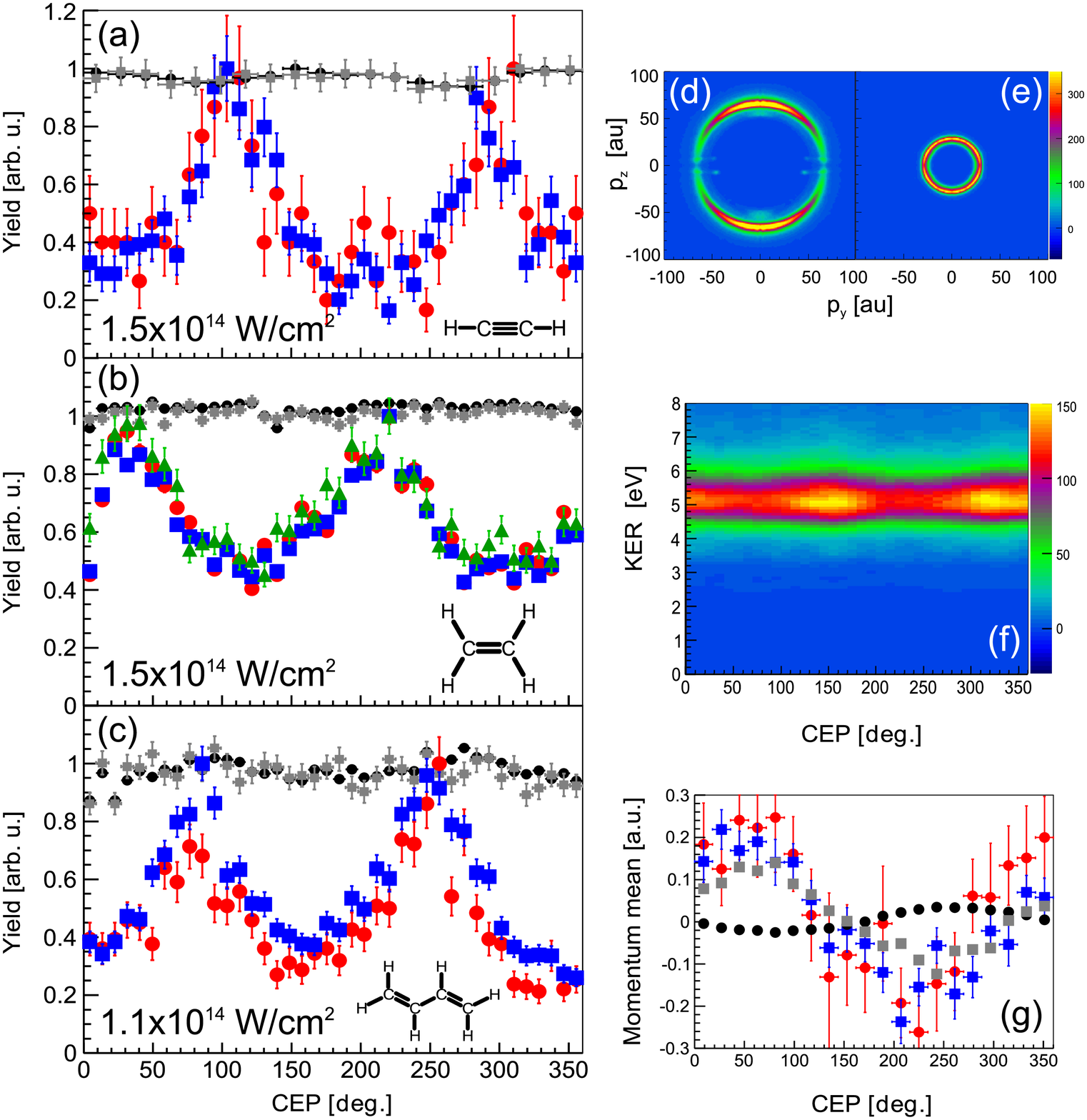}
\caption{%
(a)-(c) Measured fragmentation and ionization yields, normalized to 1 at their respective maxima, as a function of CEP for different fragmentation channels of acetylene (a), ethylene (b), and 1,3-butadiene (c), measured at the laser intensities indicated in the panels. The various fragmentation reactions are C$_2$H$_2^{2+} \rightarrow$ CH$^+ + $ CH$^+$ (red dots) and C$_2$H$_2^{2+} \rightarrow$ H$^+ + $ C$_2$H$^+$ (blue squares) for acetylene (a); C$_2$H$_4^{2+} \rightarrow$ CH$_2^+ + $ CH$_2^+$ (red dots), C$_2$H$_4^{2+} \rightarrow$ H$^+ + $ C$_2$H$_3^+$ (blue squares), and C$_2$H$_4^{2+} \rightarrow$ C$_2$H$_2^+ + $ H$_2^+$ (green triangles) for ethylene (b); C$_4$H$_6^{2+} \rightarrow$ C$_2$H$_3^+ + $ C$_2$H$_3^+$ (red dots) and C$_4$H$_6^{2+} \rightarrow$ CH$_3^+ + $ C$_3$H$_3^+$ (blue squares) for 1,3-butadiene (c). The ionization yields of the singly and doubly charged molecular ions are denoted by black dots and gray squares, respectively. 
(d,e) Cuts through the measured three-dimensional momentum spectra of CH$_2^+$ (d) and H$_2^+$ (e) associated with two of the three fragmentation channels identified for ethylene shown in (b), in a plane along ($p_z$) and perpendicular ($p_y$) the laser polarization direction, for $|p_x|<10$\,a.u. and integrated over all CEP values. 
(f) Kinetic energy release (KER) spectrum of the fragmentation reaction C$_2$H$_4^{2+} \rightarrow$ CH$_2^+ + $ CH$_2^+$ as a function of CEP.
(g) Mean $p_z$ value (integrated over $p_x$ and $p_y$) of the singly (black dots) and doubly (gray squares) charged molecular ions of acetylene as a function of CEP, in comparison with the $p_z$-sum of the two fragments from the two fragmentation channels shown in the same color and point style as in (a).
}\label{fig:yield_modulation}
\end{figure}

The main experimental results are summarized in Fig. \ref{fig:yield_modulation}. Panels (a)-(c) show the ion yields of different two-body fragmentation channels from the doubly charged ions of acetylene, ethylene and 1,3-butadiene as a function of the CEP, $\cephase$ (which exhibits a phase offset with respect to the absolute CEP, $\varphi$; see Supplemental Material  \cite{SOM}). 
The yields of all identified fragmentation channels in Fig. \ref{fig:yield_modulation} exhibit an extraordinarily strong dependence on the CEP with the strongest modulation observed for acetylene (close to 80\%). 
%
%
The fragmentation yields in Figs. \ref{fig:yield_modulation}(a)-(c) are the integrals over the three-dimensional momentum spectra of a certain fragmentation channel, uniquely identified by momentum conservation (see Supplemental Material  \cite{SOM}). Figs. \ref{fig:yield_modulation}(d, e) show examples of momentum spectra for ethylene, integrated over all CEP values. The mean momenta of the fragments show only a slight dependence on the CEP. 
The yields of the singly and doubly charged molecular ions, shown in Figs. \ref{fig:yield_modulation}(a)-(c), are also widely independent of the CEP.
The fragmentation yield, in contrast, is strongly modulated by the CEP, as can be seen from the kinetic energy release spectra, exemplary shown for the fragmentation channel C$_2$H$_4^{2+} \rightarrow$ CH$_2^+ + $ CH$_2^+$ in Fig. \ref{fig:yield_modulation}(f).


\begin{figure}[!tb]
\centering
\includegraphics[width=\columnwidth]{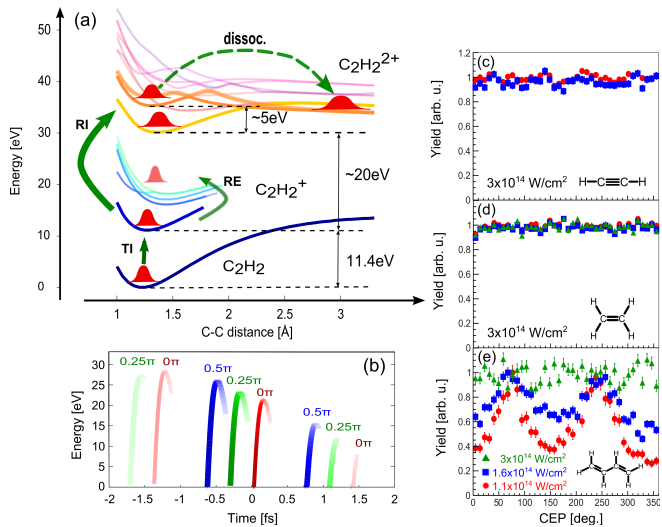}
\caption{%
(a) Electronic energy levels of the neutral, cation and dication of acetylene as a function of C-C distance, calculated by \textit{ab initio} quantum-chemical methods as described in the Supplemental Material \cite{SOM}. TI, RI, and RE denote tunnel ionization, recollision ionization and recollision excitation, respectively. 
(b) Electron recollision energy over the time of ionization calculated for a 4.5 fs (FWHM) long laser pulse with an intensity of $1.5 \times 10^{14}$ W/cm$^2$ for the three indicated values of the CEP: 0 (red lines), $0.25\pi$ (green lines), and $0.5\pi$ (blue lines). The contribution of each recollision event as determined by the ionization probability is indicated by the color intensity. 
(c, d) Fragmentation yields over CEP of the same channels as in Figs. \ref{fig:yield_modulation}(a) and (b) (same color and point styles apply), but measured for a slightly higher intensity (as indicated).
(e) Measured intensity dependence of the yield of the fragmentation channel C$_4$H$_6^{2+} \rightarrow$ CH$_3^+ + $ C$_3$H$_3^+$ over CEP. The intensities are indicated in the figure. 
}\label{fig:rec_proof}
\end{figure}


Which light-field dependent mechanism can lead to such a strong CEP dependence of the fragmentation yield?
We will discuss this question for the example of acetylene, for which the quantum chemically calculated energy level diagram for the C-C stretch mode is shown in Fig.\,\ref{fig:rec_proof}(a). Details about the quantum chemical methods for calculating the energy levels are given in the Supplemental Material  \cite{SOM}.
The energy level diagram shows that the dicationic ground state is bound. The observed fragmentation reactions, thus, have to take place on one (or more) excited states. The first dissociative state lies about 5\,eV above the dicationic ground state and can lead to both fragmentation channels observed for acetylene. Hence, to observe the strong modulation of the fragmentation yield shown in Fig. \ref{fig:yield_modulation}, the population of the involved excited state(s) has to be accomplished by a strongly CEP dependent mechanism.


As the probability for field ionization depends exponentially on the field strength and, therewith, is sensitive to the CEP of few-cycle pulses, simple sequential double ionization could be a candidate for this mechanism. Indeed, the fragmentation channels observed here have also been observed in other experiments with similarly low laser pulse intensities, but longer pulse durations ($\approx 25-50$ fs), e.g., for 1,3-butadiene \cite{Xu2010_buta}. 
In the present experiments, owing to the short duration and low intensity of the laser pulses, and supported by previous measurements with longer pulses \cite{Cornaggia2000}, we expect double ionization to occur predominantly via electron recollision. Indeed, Fig. \ref{fig:yield_modulation}(g) shows that (i) the measured momentum of C$_2$H$_2^{2+}$ along the laser polarization direction is much larger than the one of C$_2$H$_2^{+}$, and that (ii) the two momenta point into opposite directions for almost all CEP-values. 
Such CEP-dependence has been shown to be caused by double ionization via electron recollision \cite{Bergues2012}.
Furthermore, as the sum momentum of the two emitted fragments
shows the same dependence on the CEP as C$_2$H$_2^{2+}$ [see Fig. \ref{fig:yield_modulation}(g)], the fragmentation reactions are also initiated by recollision ionization (RI).
We thus conclude that the observed CEP-dependence of the fragmentation yields is most likely due to a strong CEP-dependence of RI into the first dissociative state.


\begin{figure}[!tb]
\centering
\includegraphics[width=\columnwidth]{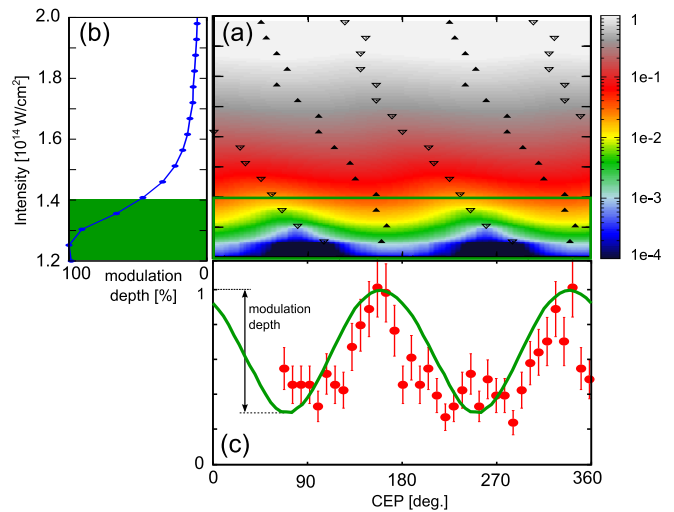}
\caption{Comparison of simulated and measured fragmentation probability of the channel C$_2$H$_2^{2+} \rightarrow$ CH$^+ + $ CH$^+$. %
(a) Simulated fragmentation probability encoded in color-scale as a function of CEP and laser intensity, normalized to its maximum value and calculated as described in the Supplemental Material \cite{SOM}. The CEP and intensity dependent position of maximum and minimum fragmentation yield is indicated by black upwards and gray downwards pointing triangles, respectively.
(b) Modulation depth of the fragmentation probability [as defined in (c)], as a function of laser intensity. 
(c) Simulated fragmentation probability over CEP (full green line) obtained by averaging of the data in (a) over the intensity range $1.2-1.4 \times 10^{14}$\,W/cm$^2$ to account for the experimental intensity beam profile [as indicated in (a) and (b)], in comparison with the experimental data from Fig. \ref{fig:yield_modulation}(a) shifted by $60^\circ$. 
}\label{fig:comp_sim_exp}
\end{figure}


In order to elucidate this point we plot in Fig.\,\ref{fig:rec_proof}(b) the electron recollision energy in the few-cycle pulses used in our experiments on acetylene, for different CEP-values as a function of birth time, calculated using the simple man's model \cite{Paulus1994, Lewenstein1995}. For this intensity ($1.5 \times 10^{14}$\,W/cm$^2$), the strongly CEP-dependent maximum recollision energy of the dominant recollision event is only 22\,eV for $\cephase = 0$, while for $\cephase = \pi/2$ roughly 26\,eV are reached. About 20\,eV are necessary to reach the ground state of the acetylene dication, see Fig.\,\ref{fig:rec_proof}(a). Additional $\sim 5$\,eV are necessary to reach the first dissociative state that leads to the experimentally observed correlated fragments CH$^+$/CH$^+$ and C$_2$H$^+$/H$^+$. 
Thus, the probability for reaching the dissociative state is highest for CEP-values around $\pi/2$, and small for CE-phases around $0$. We therefore propose the following scenario [see Fig.\,\ref{fig:rec_proof}(a)] for explaining the pronounced fragmentation yield modulations observed in the experiment:
An electronic wavepacket set free by field ionization dominantly from the HOMO of the neutral molecule, is driven back by the laser field and doubly ionizes the molecule by electron impact ionization. For CEP-values that result in low recollision energy, the second electron will be dominantly removed from the HOMO, leading to ionization to the non-dissociative dication's ground state. For CEP-values resulting in high recollision energy, the electron can also be removed from a lower lying orbital, which puts the molecular ion into a dissociative excited electronic state. For example, the excited state associated with the fragmentation products CH$^+$/CH$^+$ that we discussed above, is reached by removal of an electron from the HOMO-1 orbital. 
Because the fragmentation reactions that we investigated in the present experiments take place on only one electronically excited state surface (see Supplemental Material \cite{SOM}), whose population is controlled by the CEP, we observe the same CEP dependence for each fragmentation channel of a certain molecule (Fig.\,\ref{fig:yield_modulation}).


We can test the proposed scenario experimentally simply by increasing the laser peak intensity $I$, since the recollision energy scales linearly with intensity as $E_\textrm{rec}\propto I/4\omega^2$ \cite{Paulus1994, Lewenstein1995}.
If the recollision energy is high enough for all CEP-values to reach the first dissociative state, the fragmentation yield is expected to be independent of the CEP.
Figs.\,\ref{fig:rec_proof}(c) and (d) show the measured yield as a function of CEP of the same fragmentation channels as in Figs.\,\ref{fig:yield_modulation}(a) and (b), but with a twice higher laser intensity leading to twice the maximum recollision energy. 
As expected, the fragmentation yields now show no discernible modulation with the CEP.
Thus, by precise adjustment of the laser intensity, such that the energy threshold to the dissociative states is overcome only for certain CEP-values, light-field control of the fragmentation of polyatomic molecules becomes possible. This is demonstrated in Fig.\,\ref{fig:rec_proof}(e), for the example of the fragmentation channel C$_4$H$_6^{2+} \rightarrow $ CH$_3^+ +$ C$_3$H$_3^+$, where the laser intensity has been fine-tuned in three steps over the dissociation energy threshold, resulting first in a decreased modulation depth of the fragmentation yield, and for even higher intensity in its complete disappearance.



To obtain detailed insight into the light-determined fragmentation dynamics, we now turn to simulations (detailed in the Supplemental Material \cite{SOM}), performed exemplary for acetylene, C$_2$H$_2^{2+}$. 
Briefly, we use a one-dimensional semi-classical model to simulate recollision-ionization into the excited dissociative state.
The resulting fragmentation probability as a function of CEP and laser intensity is shown in Fig. \ref{fig:comp_sim_exp}(a).
At very low laser intensities below roughly $1.2 \times 10^{14}$ W/cm$^2$, the recollision energy of the electron is too small to reach the dissociative state, independent of the CEP. For slightly higher intensities, lower lying electrons can be ionized by RI for certain values of the CEP. With increasing intensity the CEP values, where maximum fragmentation probability is reached, shift. At the same time, the modulation depth decreases [see Fig. \ref{fig:comp_sim_exp}(b)], since for higher laser intensity and therewith higher electron recollision energy also recollisions for other CEP values have enough energy to ionize a low lying electron. 
Taking an average over $1.2-1.4 \times 10^{14}$ W/cm$^2$, thereby simulating the different intensities within the experimental laser beam profile up to an intensity very close to the experimentally determined value, we obtain the modulation shown in Fig. \ref{fig:comp_sim_exp}(c), which agrees almost perfectly with the experimental data shifted by $\sim 60^\circ$.
The CEP-shift is attributed to the influence of the long-range Coulomb field on the CEP-dependent energy of electrons from single ionization \cite{Chelkowski2004, bandrauk2005}, which we have used in our experiment to arbitrarily calibrate the CEP, $\cephase$, and which is also not correctly reproduced in the one-dimensional semi-classical model.


The numerical data shown in Fig. \ref{fig:comp_sim_exp} have been calculated by neglecting field-driven population transfer between the binding ground state and the dissociative excited states, which in previous work on the dissociative ionization of D$_2$/H$_2$ \cite{Kling2006, Grafe2007, Kling2008, Znakovskaya2012} has been shown to be essential in explaining the CEP-dependent charge-localization observed in the experiments.
Here, we can exclude a noticeable contribution of such a mechanism to the experimentally observed CEP-dependence of the fragmentation yield shown in Fig. \ref{fig:yield_modulation}, as it would be incompatible with the observation that for each of the molecules all fragmentation channels show the same CEP-dependence. Field-driven population dynamics, however, would happen on considerably different timescales for breakage of the C-C and H-C bonds, respectively, due to the large difference in the fragments' reduced masses ($\text{m}_\text{CC} / \text{m}_\text{HC} \approx 6.8$). In order to confirm this statement, we have performed reduced-dimensional quantum dynamical calculations by solving the time-dependent nuclear Schr\"odinger equation for the four strongest dipole-coupled electronic states of acetylene subjected to the experimental laser parameters (see Supplemental Material  \cite{SOM} for details).
The contribution of the field-driven population transfer to the CEP-dependent modulation of the fragmentation yield due to recollision-induced ionization was found to be very small ($< 1\%$).


Electron recollision, however, can not only promote the second electron directly into the continuum, but can also excite the molecular cation, which may then be further ionized by the laser field \cite{Kopold2000, Feuerstein2001, Rudenko2004}. 
We anticipate that this mechanism, dubbed recollision-induced excitation (RE) and subsequent field ionization (RESI), can, in principle, also be used for CEP-control of polyatomic molecular fragmentation with few-cycle pulses even at peak intensities that prohibit direct population of the first excited dicationic state. 
In this case, however, as many of the closely spaced higher excited states in the cation can be populated by RE, the CEP-selectivity becomes dominantly dependent on the field-sensitive ionization rate during the field ionization step. 
As a result, the yields of different fragmentation channels 
as well as the yield of the dication 
are likely to show different CEP-modulation depths and possibly different CEP-dependence. This is not observed in our experiments.
Furthermore, we have estimated that at the intensity of our experiments and the channels observed, RESI is considerably less probable than direct RI to the first dissociative excited state.
We therefore conclude that the contributions to the observed CEP-modulation of the fragmentation yield due to RESI, while probably important for lower intensities, are insignificant for the here demonstrated higher intensity case that permits direct RI from lower lying orbitals.


In conclusion we have experimentally demonstrated for the first time control over various fragmentation reactions of polyatomic molecules using intense few-cycle laser pulses. 
%
%
Fragmentation reactions of polyatomic molecules are essential building blocks of Chemistry. Equally important, however, are preceding isomerization reactions. In our experiment we observe, for example, two channels that involve molecular restructuring prior to the breakage of a molecular bond: formation of H$_2^+$ for ethylene and a proton migration reaction for 1,3-butadiene [see Fig. \ref{fig:yield_modulation}]. 
These results demonstrate that RI with intensity-tuned few-cycle laser pulses can be used as a very efficient tool not only to initiate (or suppress) fragmentation but also isomerization reactions in polyatomic molecules with high sensitivity. 
Our work, thus, demonstrates an efficient and selective, yet straightforward way of pre-determining fragmentation and isomerization reactions in polyatomic molecules on sub-femtosecond time-scales.


This work was partly financed by the Austrian Science Fund (FWF), grants P21463-N22, V193-N16, and I274-N16, by a SIRG grant from the ERC, and by a grant from the National Natural Science Foundation of China (No. 11074098).


\clearpage

\setlength{\textwidth}{16cm}     
\setlength{\oddsidemargin}{0cm}   
\setlength{\evensidemargin}{0cm}  
\setlength{\textheight}{22cm}   
\setlength{\topmargin}{0cm}       
\setlength{\headheight}{0cm}      
\setlength{\headsep}{0cm}         
\setlength{\footskip}{1cm}       

\renewcommand{\thefigure}{\arabic{figure}-SM}

\onecolumngrid

\begin{center}
{\bf \sffamily \huge \color{somtitlecolor} Supplemental Material}
\end{center}



\section{Experimental details}

Sub-5\,fs laser pulses were generated by spectral broadening of $\approx 25$\,fs (FWHM) laser pulses from a Titanium-Sapphire laser amplifier system, running at 5\,kHz repetition rate, in a 1\,m long hollow-core glass capillary, filled with neon, and subsequent recompression by several bounces from chirped mirrors.
The electric field of the laser pulses can be written as $E(t)=\mathcal{E}(t)\cos(\omega t + \varphi)$, where $\mathcal{E}(t)$ is the pulse envelope and $\omega$ and $\varphi$ are the light frequency and the carrier-envelope phase (CEP), respectively.
The few-cycle pulses, linearly polarized along the $z$-direction of our lab coordinate system, with their spectra centered around 750\,nm, are directed into an ultra-high vacuum chamber (background pressure $1.3 \times 10^{-10}$\,mbar), where they are focused onto a supersonic molecular gas jet of randomly aligned acetylene (C$_2$H$_2$), ethylene (C$_2$H$_4$), or 1,3-butadiene (C$_4$H$_6$) molecules by a spherical mirror with a focal length of 150\,mm. The gas jet, propagating along the $x$-direction, was created by expanding the molecular gas samples, with a backing pressure of around 0.5\,bar (depending on the molecular species), through a nozzle of 10\,$\mu$m in diameter into a vacuum chamber at $\sim 4 \times 10^{-5}$\,mbar operating pressure and subsequent two-stage skimming and differential pumping.
To avoid smearing of the CEP due to the Gouy phase shift when the laser pulses propagate along the $y$-direction through the gas jet \cite{Lindner2004a}, which is $\approx 170$\,$\mu$m in diameter, the jet is cut in width by an adjustable slit before it reaches the interaction region. This reduces the width of the jet to typically 30\,$\mu$m. The laser focus is placed about half the Rayleigh length before the center of the narrow jet. The Rayleigh length is estimated as 150\,$\mu$m.

Cold Target Recoil Ion Momentum Spectroscopy (COLTRIMS) \cite{Doerner2000} is used to measure the three-dimensional momentum vector of fragment ions emerging from the interaction of a single molecule with a few-cycle laser pulse.
The ions created by the laser pulses were guided over a distance of 5.7\,cm by a weak homogeneous electric field of 16\,V/cm, applied along the axis of the spectrometer, $z$, onto a multi-hit capable RoentDek DLD 80 detector (80 mm in diameter), equipped with position sensitive delay line anodes. The ion count rate was kept at 0.4 per laser shot in order to establish coincidence conditions, which ensure that all observed processes take place within a single molecule. 
From the measured time-of-flight and position of each detected ion we calculate its three-dimensional momentum vector in the lab frame \cite{Doerner2000, Zhang2012}. 
The peak intensity of the laser pulses on target was determined from separate calibration measurements using single ionization of argon atoms in circularly polarized light \cite{Maharjan2005, Smeenk2011}.

As in our experiments we measure the ion momenta of fragments from only one molecule created by ionization in a single laser pulse, it becomes possible to perform single-shot CEP tagging \cite{Johnson2011}: We measure the duration and the CEP, $\varphi$, of each few-cycle laser pulse on a single-shot basis by exploiting the asymmetry of photoelectron spectra created by above-threshold ionization (ATI), emitted to the left and right along the laser polarization direction, using a stereo-ATI phase-meter \cite{Wittmann2009, Sayler2011, Sayler2011a}.
In the offline data analysis each fragmentation event is then tagged with the corresponding CEP-value determined from the photoelectron spectra \cite{Sayler2011, Sayler2011a}. 
From the stereo-ATI data we estimate the duration of our laser pulses as $4.5$\,fs full width at half maximum (FWHM).

As CEP-measurement and coincidence measurement of fragment ions are performed in two separate apparatus, the laser beam, split into two arms, passes through different amounts of glass and air. Thus, the CEP-value determined from the stereo-ATI phase-meter, $\varphi$, is offset by a constant value, $\varphi_0$, from the one of the pulses in the COLTRIMS device, $\cephase$. 
We arbitrarily calibrated the CEP in the COLTRIMS device such that the mean momentum value of the singly charged molecular ion, which due to momentum conservation is the inverse of the electron momentum, peaks at $\cephase=90^\circ/270^\circ$ and is zero at $\cephase=0^\circ/180^\circ$, see Fig.\,1(g) in the Letter. This CEP-dependence is predicted by the simple man's model (SMM) of strong field physics \cite{Lewenstein1995}, which neglects the influence of the long-range Coulomb potential onto the departing electrons. Due to the Coulomb influence, depending on the atomic or molecular species and on the laser parameters, the CEP-values at which the maxima of the mean momentum are observed, do not necessarily coincide with the predictions of the SMM \cite{Chelkowski2004, bandrauk2005}. 
As the Coulomb influence and laser parameters are different for the separate measurements shown in the Letter, $\varphi_0$ might also be different for each of them.
We would like to emphasize, though, that all explanations and conclusions given in the Letter are unaffected by a constant CEP-offset, $\varphi_0=\varphi - \cephase$.


From the measured data a certain two-body fragmentation pathway is uniquely identified using momentum conservation conditions in all three spatial dimensions \cite{Hasegawa2001, Xu2010_buta, Zhang2012}. From all detected ions per laser pulse we select only those sets of two fragment ions, for which the absolute value of their momentum sum is smaller than 3 a.u. along the $y$ and $z$ directions, and smaller than 5 a.u. along the $x$ direction. By that we identified two different fragmentation pathways for acetylene and 1,3-butadiene, and three different ones for ethylene (see caption of Fig. 1 in the Letter).


\section{Potential energy surfaces of acetylene from \textit{ab initio} quantum chemical methods\label{sec:PES}}

\begin{figure}[htbp]
\centering\ 
\includegraphics[width=0.6\textwidth,angle=0]{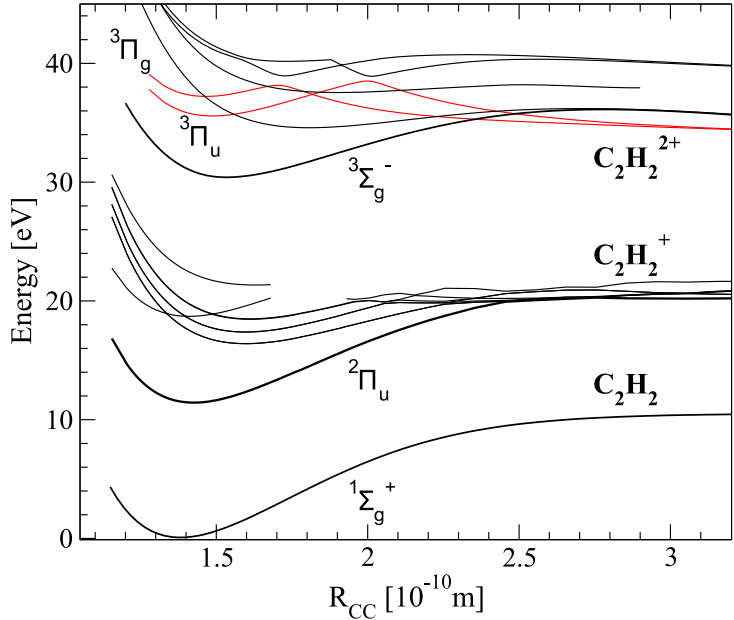}
\caption{\textbf{\textit{Ab initio} potential energy curves of C$_2$H$_2$, C$_2$H$_2^+$ and C$_2$H$_2^{2+}$.} The C-H distance, $R_\text{CH}$, is frozen at the equilibrium distance of the dication, $R_\text{CH}=1.15 \mathring{A}$.}
\label{fig:som_pes}
\end{figure}

\begin{figure}[htbp]
\centering\ 
\includegraphics[width=0.6\textwidth,angle=0]{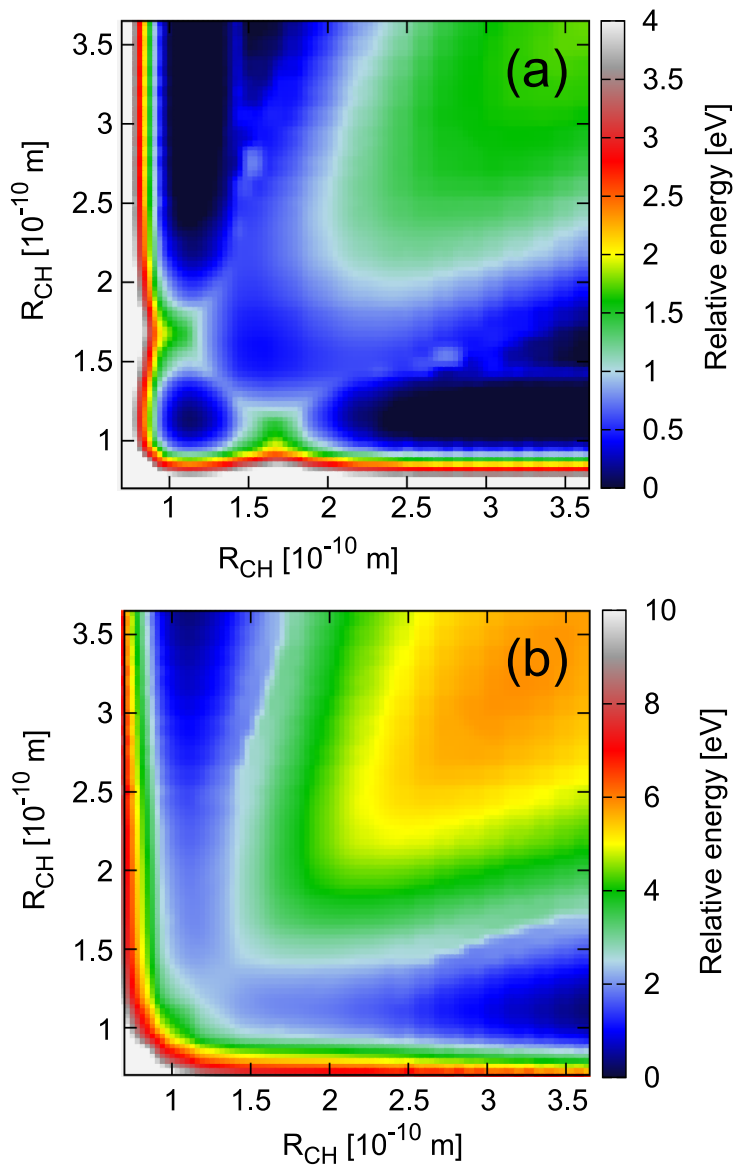}
\caption{\textbf{Calculated potential energy surfaces of C$_2$H$_2^{2+}$.} Electronic ground state (a) and first excited state ($^3\Pi$) (b) for the C-H stretch modes with the C-C distance, $R_\text{CC}$, frozen at the equilibrium distance of the dication, $R_\text{CC}=1.15 \mathring{A}$.}
\label{fig:surfaces}
\end{figure}

\begin{figure}[htbp]
\centering\ 
\includegraphics[width=0.74\textwidth,angle=0]{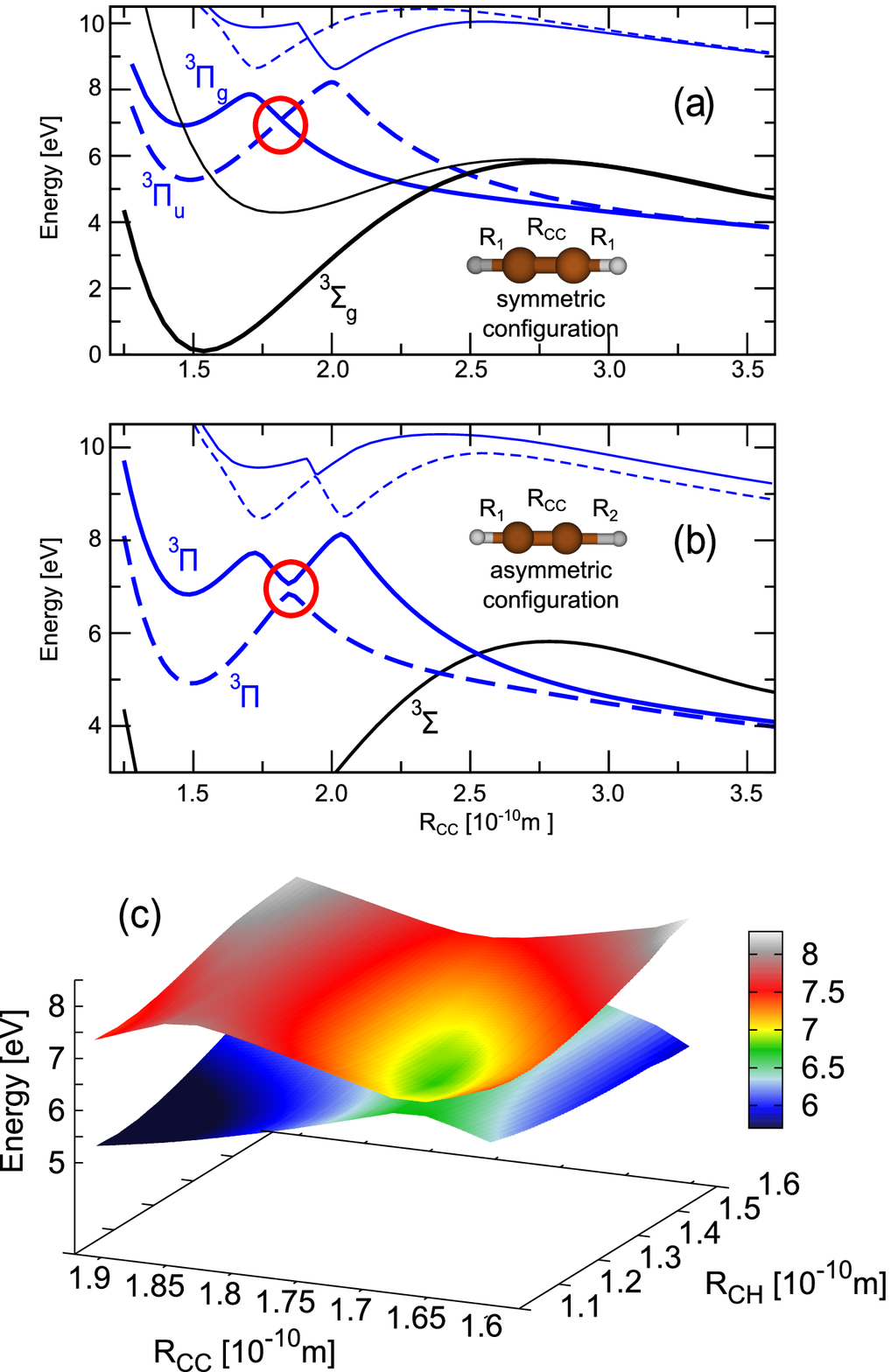}
\caption{\textbf{Selected cuts of the potential energy surfaces of C$_2$H$_2^{2+}$ for the C-C stretching mode.} Black curves: Electronic ground state and the first dipole coupled state of the dication. Blue curves: the first excited electronic states (weakly dipole coupled to the electronic ground state in the direction perpendicular to the molecular axis). (a) For a symmetric C-H distance, the $^3\Pi$-states show a crossing. (b) At a slightly asymmetric C-H/C-H distance, the two $^3\Pi$ states show an avoided crossing in 1 dimension (red circle), and the first excited state becomes more pre-dissociative. (c) Two-dimensional potential energy surfaces showing the conical intersection between the $^3\Pi$ states as a function of one C-H and C-C distance. The other C-H bond is kept frozen at $R_\text{CH}=1.15 \mathring{A}$. Comparison with Fig.\,\ref{fig:surfaces}(b) shows, that this asymmetric C-H stretch will inevitably happen, thus, opening up the fragmentation pathway.}
\label{fig:levels}
\end{figure}

The potential energy surfaces (PES) of acetylene (C$_2$H$_2$) and its ionic states (C$_2$H$_2^+$, C$_2$H$_2^{2+}$) were calculated using high-level \textit{ab initio} quantum chemical methods. The calculations were carried out using a standard correlation-consistent polarized double-zeta basis set with augmented functions (aug-ccpVDZ) \cite{Dunning1989}. For the calculations of the potential curves and surfaces, multi-reference methods (CASSCF \cite{Knowles1985} and MR-CI\cite{Werner1988,Knowles1988}) were employed, where all valence electrons of the molecule were taken into the active space, the carbon 1$s$ orbitals were kept frozen. The valence electrons were distributed in at least 8 orbitals, (8,8) CASSCF, but active spaces up to 11, (8,11) CASSCF were also examined. In the CASSCF calculations, the state-averaging procedure over the lowest five states was included. Subsequently, MR-CISD calculations with a (8,8) reference space were carried out. All calculations were performed in the frame of the C$_{2v}$ point group (with the symmetry axis being along the C-C bond) in order to be able to treat all strech vibrations examined here on the same footing. The \textit{ab initio} calcuations have been performed using the \textsc{Columbus} \cite{columbus1, columbus2, columbus3} software package.

Cuts of the PES of the neutral, singly, and doubly ionized acetylene describing stretching along the C-C bond ($R_\text{CC}$) are shown in Fig.\,2(a) in the Letter, and enlarged in Fig.~\ref{fig:som_pes}. The electronic ground state of the neutral is $^1\Sigma_g^+$, $^2\Pi_u$ for the cation, and $^3\Sigma_g^-$ for the dication. Also shown in Fig.~\ref{fig:som_pes} are some of the lowest lying doublet states of the cation, and some of the lowest lying triplet states of the dication. The vertical ionization potential is found to be 11.4~eV. In the cation, several higher lying doublet electronic states are present with (vertical) excitation energies between 5 and 9~eV above the doublet ground state (see also Ref.~\cite{Peric1998a,Peric1998b}), which may be populated optically or by electron recollisional excitation (see Letter). The triplet groundstate of the dication $^3\Sigma_g^-$ is 31.5\,eV above the neutral acetylene and is bound along the C-C, as well as along the C-H stretch motion, see Fig.~\ref{fig:surfaces}(a). 
The singlet groundstate ($^1\Delta_g$) lies about 1~eV above the energy of the triplet groundstate and is bound as well.

The first excited triplet state in the dication ($^3\Pi_u$) lies energetically about 5\,eV above the $^3\Sigma_g^-$ groundstate. This state is dissociative along the C-H stretch (and also along the $D_{\infty h}$ distorting modes \cite{Zyubina2005}), leading to the fragmentation products H$^+$/C$_2$H$^+$, see Fig.~\ref{fig:surfaces}(b). It is not immediately obvious, however, that the $^3\Pi_u$ state is also dissociative along the C-C stretching mode, since the first excited triplet state has a symmetry-allowed level crossing with the higher excited $^3\Pi_g$ state, see Fig.~\ref{fig:levels}(a). This effectively leads to the $^3\Pi_u$ state being bound along the mirror-symmetric C-C stretching mode. For non-symmetric configurations, however, the degeneracy is lifted [see Figs.~\ref{fig:levels}(b) and \ref{fig:levels}(c)] and the state crossing between the two $^3\Pi$ states becomes avoided (in one dimension). As a result, fragmentation into the products CH$^+$/CH$^+$ becomes possible. Breaking of the symmetry will inevitably happen since the state is dissociative along the antisymmetric C-H stretch motion, as well as for distortions from the linear geometry (where the $^3\Pi$ state splits up into two components with $^3 A'$ and $^3 A''$ symmetry, respectively). 
Please note, that breakage of the C-C bond may also 
proceed via different electronic states. 
In particular the $^1\Sigma_g^+$ state with a small dissociation barrier of about 3\,eV, that can be overcome for higher recollision energies, leads to the fragmentation products CH$^+$/ CH$^+$ \cite{Duflot1995}.

From this analysis we can deduce that an excitation to the first few excited states, about 25-26\,eV above the cationic groundstate, will suffice to lead to a fragmentation into both fragmentation channels observed in the experiment.


\section{Quantum dynamical simulations of field-driven fragmentation dynamics\label{sec:dynamics}}

With the \textit{ab initio} potential energy surfaces $V_i(R_1,R_2,\dots)$ and transition dipole moments $\vec\mu_{ij}(R_1,R_2,\dots)$ calculated as described in Section~\ref{sec:PES}, we have performed quantum dynamical calculations, integrating numerically the time-dependent nuclear Schr\"odinger equation in the presence of an intense, external laser field. The calculations included the lowest four electronic states, two $^3\Sigma$ states, as well as the two $^3\Pi$ states. The $^3\Sigma$ (and $^3\Pi$) states have a large transition dipole moment along the internuclear axis, while the transition dipole moment between the $\Sigma$ and $\Pi$ states of different symmetry are very small. Thus, only the coupling from the electronic ground state to the lowest lying excited electronic state, $^3\Sigma_g^- \rightarrow ^3\Pi_u$, is included. The nuclear Schr\"odinger equation reads as (atomic units are employed here and in what follows)
\begin{equation}
\label{eq:Nuc_Seq}
i\frac{\partial}{\partial t} \left(\begin{array}{c} 
\psi_1(\vec{R},t) \\ \psi_{\Pi1}(\vec{R},t) \\ \psi_{\Pi2}(\vec{R},t)\\ \psi_2(\vec{R},t) \\
\end{array} \right) = H \left(\begin{array}{c} 
\psi_1(\vec{R},t) \\ \psi_{\Pi1}(\vec{R},t) \\ \psi_{\Pi2}(\vec{R},t)\\ \psi_2(\vec{R},t) \\
\end{array} \right), 
\end{equation}
where the Hamiltonian is given by
%
%
\begin{align}
\label{eq:Hamiltonian}
H &= T + \nonumber \\
&\quad\left(\begin{array}{cccc} 
V_1(\vec{R}) & -\mu_{1,\Pi1}(\vec{R})\vec{E}(t)&0&-\mu_{1,2}(\vec{R})\vec{E}(t) \\ 
-\mu_{1,\Pi1}(\vec{R})\vec{E}(t)& V_{\Pi1}(\vec{R}) &-\mu_{\Pi1,\Pi2}(\vec{R})\vec{E}(t)&0 \\ 0&-\mu_{\Pi2,\Pi1}(\vec{R})\vec{E}(t)& V_{\Pi2}(\vec{R}) &0\\ 
-\mu_{2,1}(\vec{R})\vec{E}(t)&0&0& V_2(\vec{R}) \\
\end{array} \right).
\end{align}

In the last equation, $T$ is the kinetic energy operator, $\vec\mu_{ij}$ the transition dipole moment between states $i,j$ with components parallel and perpendicular to the molecular axis, and $\vec{E}(t)$ denotes the vectorial laser electric field. Denoting the angle between the laser polarization direction and the molecular axis with $\vartheta$, $-\mu_{ij}^{z}\cos \vartheta\, E(t)$ and $-\mu_{ij}^{\perp}\sin \vartheta \,E(t)$ then represent the projections of the interaction with the electric field onto the directions along and perpendicular to the molecular axis, respectively. In Eq.~\ref{eq:Hamiltonian}, $\vec{R}$ represents one or more selected vibrational coordinates, either the C-C or the C-H vibrations. The reduced nuclear mass $M_{red}= (M_a M_b)/(M_a +M_b) $ contains the nuclear masses $M_{a,b}$ of the atoms $a,b$ of the selected degree-of-freedom. The time-dependent nuclear Schr\"odinger equation is solved numerically on a grid using the split-operator method \cite{Feit1982}. 

For the laser field, we applied a few-cycle pulse with a wavelength of 800~nm and an intensity of $1.4 \times 10^{14}$ W/cm$^2$ with a duration of 5~fs (FWHM). Different angles $\vartheta$ of the field relative to the molecular axis have been simulated. The initial conditions were also varied, mimicking either (i) initial nuclear wavepackets in field-dressed states or, (ii) nuclear wavepackets with an initial population ratio of the ground and excited state as given by the recollisional ionization estimate described in Section~\ref{sec:model}. The population dynamics $P_i(t) = \int |\psi_i(R,t)|^2 dR$ features pronounced oscillations following closely the oscillations of the laser field, with maximum values of the population of the excited state, $P_{2}$, up to 0.025. The oscillations in the population dynamics of the $\Pi$ states, $P_{\Pi1}, P_{\Pi2}$, although following the oscillations of the laser field as well, are much less pronounced and reach maximum values in the population of about 10 times weaker than the values found for the state labeled with 2. 

The final population of the $^3\Pi$ state(s) was almost independent of the CEP for all angles $\vartheta$ and all examined pulses with different pulse lengths and intensities. Population transfer in the dication induced by the laser field seems to be of minor importance to describe the large differences in the fragmentation yield as a function of CEP observed in the experiment. Hence, only the different initial population of the $\Pi$ state being populated by electron recollisional ionization of the cation can explain the large variation in the fragmentation yield -- additional field excitation in the dication only gives a minor contribution.


\section{Modeling of CEP-dependent fragmentation via recollision-ionization into the excited dissociative state\label{sec:model}}

To simulate the mechanism for CEP-control of fragmentation based on recollision-ionization (as outlined in the Letter), we use a one-dimensional semi-classical model. 
We consider fragmentations into C$_2$H$^+$/H$^+$ (H-C stretching mode) and CH$^+$/CH$^+$ (C-C stretching mode) that were observed in the experiment. As detailed in Section~\ref{sec:PES}, the first states leading to a strong dissociation into C$_2$H$^+$/H$^+$ and CH$^+$/CH$^+$ lie $\sim 25$\,eV above the ionic groundstate, and population of these states will lead to a fragmentation into both channels.
In the model we assume that the first electron is set free by field ionization, with an ionization rate according to tunneling theory \cite{Ammosov1986},
\begin{equation}
\label{eq:adk}
\Gamma = \text{const} \cdot \left(\frac{2 (2 I_p)^{3/2}}{|E(t_b)|}\right)^{2n-|m|-1}\,\exp\left(-\frac{2 (2 I_p)^{3/2}}{|E(t_b)|}\right),
\end{equation}
for an ionization potential of $I_p = 11.4$\,eV and quantum numbers $l = m = 1$, as suggested in Ref.~\cite{Tong2002} for ionization from $\pi$-bonds. The electron is placed at a distance $z_0=I_p/E(t_b)$, with $E(t_b)$ the electric field at birth time $t_b$, from the nucleus with a Gaussian momentum distribution parallel to the laser polarization direction, $p_z$, according to tunneling theory \cite{krainov1997}. Neglecting the initial longitudinal velocity distribution, however, did not change the results markedly. Wavepacket spreading is estimated via a quadratic increase of its size $A$ perpendicular to the laser polarization direction with time \cite{Delone1991}:
\begin{equation}
A=\pi\frac{|E|}{2\sqrt{2I_p}}(t-t_b)^2.
\end{equation}
The electron, after its birth at time $t_b$, is propagated classically in the laser field and a soft core Coulomb potential $V(z) = -1/\sqrt{(1 + z^2)} $ by numerically solving Newton's equations. 
The electron is defined as recolliding with the molecule, if it returns to the ion's position. Only the first recollision event is taken into account. The corresponding recollision energies for different values of the CEP are displayed in Fig.\,2(b) of the Letter. Clearly, the recollision energy features a strong dependence on the CEP. 

The probability for recollision-ionization (RI) to the excited ionic states is derived from the energy dependent impact ionization cross-section $\sigma$ \cite{medvedev2010}, scaled by $\sigma/A$. The cross-section is zero below 25\,eV, the minimum excitation energy to the first dissociative state, followed by an approximately linear rise as a function of energy \cite{medvedev2010}. Arbitrary scaling factors of $\sigma$ are unimportant since we are only interested in relative changes over CEP. The resulting fragmentation probability as a function of CEP and laser intensity is shown in Fig.\,3(a) of the Letter.


\vspace{1cm}

\twocolumngrid


%

\end{document}